# Ballistic Josephson junctions based on CVD graphene

Tianyi Li,[1,2,3] John Gallop,[3] Ling Hao,[3] and Edward Romans[1,2]

[1]London Centre for Nanotechnology, University College London, London WC1H 0AH, UK

[2]Department of Electronic and Electrical Engineering, University College London, London WC1E 6BT, UK

[3]National Physical Laboratory, Teddington TW11 0LW, UK

## Abstract

Josephson junctions with graphene as the weak link between superconductors have been intensely studied in recent years, with respect to both fundamental physics and potential applications. However, most of the previous work was based on mechanically exfoliated graphene, which is not compatible with mass production. Here we present our research using graphene grown by chemical vapour deposition (CVD) as the weak link of Josephson junctions. We demonstrate that CVD-graphene-based Josephson junctions with Nb electrodes can work effectively without any thermal hysteresis from 1.5 K down to a base temperature of 320 mK, and they show an ideal Fraunhofer-like interference pattern in a perpendicular magnetic field. We also show that the critical current of the junction can be tuned by a gate voltage. Furthermore, for our shortest junctions (50 nm in length), we find that the normal state resistance oscillates with the gate voltage, indicating that the junctions are in the ballistic regime, a feature not previously observed in CVD-graphene-based Josephson junctions.

The experimental isolation of mono-layer graphene[1] has triggered an enormous effort to explore the new physical properties of graphene-superconductor hybrid structures. The unique band structure of graphene led to the prediction in 2006 that superconductor-graphene-superconductor (SGS) Josephson junctions should have a bipolar critical current that can be tuned from the electron band to the hole band by application of a gate voltage.[2] Such a unique property introduces a new degree of freedom for device operation, and offers the potential for a range of novel applications in magnetic metrology, quantum information, and other graphene-based electronics. Since the first SGS junction was experimentally realised by Heersche et al. in 2007,[3] a number of papers on graphene- or thin-graphite-based Josephson junctions and SQUIDs have been published,[4–25] and many novel phenomena including multiple Andreev reflection[3,4,6–8,14,15,18–20,23] and ballistic transport[15,17–20,23] have been discovered and discussed. The earlier work on SGS junctions was mainly based on mechanically exfoliated graphene,[3–14,16,21,22] due to the high quality and easy availability of such samples. To further investigate the new physics in the ballistic regime of the SGS junctions, various treatments had been made on mechanically exfoliated graphene, such as suspending it,[15,23] or encapsulating it in hexagonal boron nitride (hBN).[17,20] Recently, a few reports on SGS junctions based on epitaxial graphene on a SiC substrate[24] or graphene grown by CVD[25] have been published. However, the fabrication and measurement of SGS junctions compatible with large scale production remain less explored, and applications involving multiple SGS junctions are still a long way off.

Here we report our study on SGS junctions based on CVD graphene. The fabrication process has been carefully optimised to be compatible with very short and wide junctions. We show that these junctions exhibit ideal *I-V* characteristics and an ideal Fraunhofer pattern in a perpendicular magnetic field. We also show that the critical current of the junction can be effectively tuned by the gate voltage. Conduction through junctions based on CVD graphene is usually assumed to be diffusive due to a short mean free path.[25] However, our carefully controlled fabrication technique allows us to build junctions as short as 50 nm, which appears to be shorter than the mean free path. Consequently we managed to observe the oscillation of the junction normal state resistance versus gate voltage in more than one device, indicating that our shortest junctions are in the ballistic regime. The ballistic transport is also consistent with the behaviour of the critical current as a function of temperature.

We have developed a robust process to fabricate Josephson junctions based on CVD graphene, which is compatible with mass production. The graphene samples used in our experiment (Graphenea™) were synthesised and transferred to a silicon substrate with a 300-nm-thick oxide layer. The transferred graphene was initially characterised by atomic force microscopy (AFM) and Kelvin probe force microscopy (KPFM), which showed that the silicon substrate was fully covered by graphene, mostly mono-layer (see Supporting Information for the images). After that, we use electron beam lithography (EBL) to define the superconducting electrodes. In order to reduce the junction normal state resistance $R_n$ and thus to increase the junction critical current $I_c$, the Josephson junctions we define are of short length $L$, in the range 50-250 nm, and large width $W = 80$ μm, i.e. with a typical aspect ratio of ~1/1000 (Figure 1). After developing the resist, a tri-layer of Ti (5 nm)/Nb (70 nm)/Au (8 nm) is sputtered onto the sample, followed by lift-off without sonication. Ti is used here as an adhesive layer, while Au is sputtered immediately after Nb to prevent its oxidation. Since sputtered atoms are less directional than evaporated atoms, sputtering and lift-off are somewhat incompatible with each other in a normal micro-fabrication process, especially for small structures with dimensions of tens of nanometres. However, by using double-layer PMMA resist and sputtering the tri-layer at a lower rate to prevent overheating the sample, we managed to fabricate short and wide Josephson junctions with a very high throughput. We have inspected the fabrication of hundreds of devices in a single EBL process, and the visible yield is higher than 95%. After defining the junctions, another EBL process is used to remove the graphene surrounding the devices by Ar milling, so that the devices are isolated from one another and the junctions have no additional shunt resistance outside the gap between the electrodes. The highly-doped silicon substrate is connected as the back gate electrode. (See Supporting Information for fabrication details.)

We measured the electronic properties of the graphene-based Josephson junctions using a normal 4-probe configuration, in a $^3$He cryostat with a base temperature of 320 mK. As shown in Figure 2(a), the Josephson junctions exhibit normal *I-V* characteristics at the base temperature without any hysteresis, as predicted by the resistively shunted junction (RSJ) model for over-damped junctions. As temperature gradually increases, the critical current $I_c$ becomes smaller, and the transition close to $I_c$ becomes slightly rounded, which is the common behaviour of a strongly overdamped junction with increasing thermal noise. The Josephson effect is still obvious up to 1.50 K, indicating that the junction can operate over a wide temperature range. By plotting the differential resistance *dV/dI* as a function of bias current (Figure 2(b)), we can see the trend more clearly. The two peaks in the *dV/dI* curves, indicating the transition near $I_c$, gradually become weaker in height and closer to each other as the temperature increases.

To investigate this further, we plot the critical current as a function of temperature for junctions with lengths of 50 nm, 150 nm and 250 nm, in Figures 2(c)-(e) respectively. For each device, the critical current $I_c$ increases as the temperature $T$ decreases. However, the $I_c$-$T$ curves are not identical in shape. For the 150-nm- and 250-nm-long junctions, the critical current $I_c$ tends to saturate when the temperature is below a certain value, whereas for the 50-nm-long junction, the critical current $I_c$ is still increasing at the lowest temperatures. Such a difference in the $I_c$-$T$ curves can be well explained by the theory of weak-links in the "dirty" or "clean" limit developed by Kulik and Omelyanchuk (the KO-1 and KO-2 theories).[26,27] A weak-link junction is in the dirty limit if the junction length $L$ is much longer than the mean free path $l$, and is in the clean limit if $L \ll l$. For a weak-link junction in the dirty limit, as shown by the KO-1 theory, the supercurrent $I_s$ is related to temperature $T$ by

$$I_s(T,\varphi) = \frac{2\pi k_B T}{eR_n} \sum_{\omega>0} \frac{2\Delta \cos(\varphi/2)}{\delta} \arctan \frac{\Delta \sin(\varphi/2)}{\delta} \quad (1)$$

where $R_n$ is the normal state resistance of the junction, $\Delta$ is the temperature dependent energy gap of the superconductor, $\omega = \pi k_B T(2n+1)/\hbar$ is the Matsubara frequency for integer $n$, $\varphi$ is the phase difference across the weak-link, and $\delta = \sqrt{\Delta^2 \cos^2(\varphi/2) + (\hbar\omega)^2}$. For a weak link in the clean limit, the $I_s$-$T$ relationship is described by the KO-2 theory,

$$I_s(T,\varphi) = \frac{\pi\Delta}{eR_n} \sin(\varphi/2) \tanh \frac{\Delta \cos(\varphi/2)}{2k_B T} \quad (2)$$

For a given $T$, the expressions in Eqs. (1) and (2) need to be maximised over $\varphi$ to determine $I_c(T)$. Qualitatively speaking, the $I_c$-$T$ curves determined by the KO-1 and KO-2 theory almost coincide with each other when the temperature is close to $T_c$.[28] As temperature decreases, the $I_c$ of a dirty junction tends to saturate while the $I_c$ of a clean junction is still increasing. As shown by the red curves in Figures 2(c)-(e), the measured $I_c$-$T$ curves of the 150-nm- and 250-nm-long junction can be well fitted by the KO-1 theory, while the $I_c$-$T$ curves of the 50-nm-long junction is better fitted by the KO-2 theory (see Supporting Information for detailed fitting results). This suggests that junctions longer than 150 nm are in the dirty regime where the transport of charge carriers is diffusive, while the 50-nm-long junction is in the clean regime where ballistic transport takes place. According to the fits, the junction transition temperature $T_c$ is 1.51~1.73 K, which is much smaller than that of the superconducting electrodes ($T_c \sim 9$ K). This is probably because the interface between the graphene and the superconducting electrode is not highly transparent, so that the superconducting gap induced in the graphene by the proximity effect is much weakened. We also note that the critical current $I_c$ at the base temperature does not show a strong dependence on the junction length. This is because the contact resistance between the graphene and the superconducting electrode, which shows some device-to-device variation, makes a large contribution to the normal state resistance, as will be confirmed below. As a result, the normal state resistance $R_n$ also shows some device-to-device variation and is not so sensitive to the junction length. Since $I_c$ is inversely proportional to $R_n$ in both the KO-1 and KO-2 theories, $I_c$ does not show a strong dependence on the junction length for the range of lengths we have considered.

We also measured the electronic properties of the junctions in a perpendicular applied magnetic field. As Josephson junctions with finite area, they are expected to show the effect of quantum interference, since the perpendicular magnetic field can induce a phase difference between the supercurrent at different points across the junction width. In Figure 3, we plot $dV/dI$ as a function of bias current $I$ and magnetic field $B$ in a 2D colour scale plot. The purple region is where the differential resistance is zero, and the outline of the purple region indicates the critical current. As can be seen in the plot, the outline of the purple region can be perfectly fitted by an ideal Fraunhofer-like pattern, $I_c \propto \sin(\pi B/\Delta B)/(B/\Delta B)$, as indicated by the black dashed lines, where $\Delta B$ is the field needed to reach the first minimum. The higher-order peaks, though not as high in contrast as the central ones, are still clearly visible. In another measurement on the same device, we managed to see peaks up to $\pm 80$ µT (Figure S3 in Supporting Information). The observation of the Fraunhofer pattern indicates that the junction, although having an aspect ratio of 1/320, is still quite uniform in terms of the distribution of its supercurrent density. Since the first minimum corresponds to one flux quantum $\Phi_0$ in the junction, we can calculate the effective area of the junction as $A_{eff} = \Phi_0/\Delta B = 280$ µm$^2$. This is much larger than the effective area naively estimated as $A'_{eff} = (L+2\lambda)W = 26$ µm$^2$, corresponding to the area of the normal region between the electrodes plus the area of the electrodes the field penetrates into, set by the penetration depth $\lambda$. However the latter estimate ignores the strong flux focussing caused by the Meissner effect in the electrodes, which each extend 5 µm away from the junction region meaning their area is much larger than that of the junction. The measured effective area corresponds to about 35% of the flux applied to each electrode being forced into the junction area, so this is a reasonable explanation for the size of the discrepancy.

One of the key advantages of a graphene-based Josephson junction is that the critical current can be effectively tuned by the gate voltage, thanks to the relatively low density of states in graphene.[3–6,8–15,18,20–22,25] In Figure 4(a) we show *I-V* curves measured on the same device under different gate voltages. Unlike the case in Figure 2(a), the gate voltage not only affects the critical current, but also the normal state resistance. The effect is more clearly demonstrated if we plot *dV/dI* as a function of bias current and gate voltage in a 2D colour plot (Figure 4(b)). As the gate voltage increases from -50 V to 50 V, the critical current (the boundary of the dark blue region) gradually decreases from 8.5 µA to 1.5 µA, i.e. a nearly 6-fold change. However, even when gate voltage is as large as 50 V, we still do not reach the Dirac point of the graphene. This is simply because our particular CVD graphene sample is heavily p-doped, and the oxide layer of the silicon substrate is too thick to effectively tune the Fermi level in the graphene. Since our fabrication technique is suited to any CVD graphene sample, this problem can be easily solved in future. Note that the discontinuity that appears at ±25 V is purely an artefact from the measurement system. Our junctions are found to be very sensitive to noise, and the source for gate voltage happens to have higher ac noise level while working in the range from 25 V to 50 V.

The product of critical current $I_c$ and normal state resistance $R_n$ is the characteristic switching voltage when the junction jumps from the superconducting state to the normal state around the critical current. For a weak-link Josephson junction in the dirty limit, the product $I_cR_n$ is a figure of particular interest since it is solely determined by the operating temperature and the energy gap of the superconductor, as indicated by the form of Eq. (1). If the temperature is much lower than $T_c$, the product should be a constant

$$I_cR_n \sim 2.07\,\Delta_0/e \tag{3}$$

where $\Delta_0 \approx 1.764 k_B T_c$ is the superconducting energy gap at $T = 0$ K. In Figure 4(c) we plot both the normal state resistance $R_n$ and the product $I_cR_n$ as a function of the gate voltage, for the 150-nm-long junction. As the gate voltage draws closer to the Dirac point, the critical current decreases while the normal state resistance increases. Their product $I_cR_n$ remains a constant when the gate voltage is far away from the Dirac point; however, it drops as the gate voltage approaches the Dirac point. Even in the constant regime, the measured value of $I_cR_n$ is only ~1/10 of the theoretical one determined by Eq. (3). Such suppression of $I_cR_n$ is frequently discussed in previous reports on graphene-based Josephson junctions, and there still remains some controversy about the exact cause. Many authors believe that the overall suppression of $I_cR_n$ is a result of the imperfect interface between graphene and the superconductor.[13,15,18,25] As to the further drop of $I_cR_n$ near the Dirac point, most authors believe that it is caused by a discrepancy between the intrinsic critical current and the measured critical current due to premature switching, which is expected to be more pronounced near the Dirac point.[3,4,6,8] However, others attribute the suppression of $I_cR_n$ to specular Andreev reflection that occurs close to the Dirac point.[13]

Figure 4(d) shows the normal state resistance of a 50-nm-long SGS junction as a function of gate voltage, which displays a similar trend as Figure 4(c). However, we can also observe small oscillations in the $R_n$-$V_g$ curve. If we fit the curve by a smooth polynomial, and plot the fitting residue versus $V_g$, the periodic oscillation can be more easily seen, as shown in the lower panel of Figure 4(d). We believe that such oscillations are due to the phase-coherent interference of charge carriers in the Fabry-Perot cavity defined by the n-p-n junction that arises from the different doping levels in the exposed graphene and the unexposed graphene (under the electrodes). Such oscillations provide more evidence that the junction is in the ballistic regime. The oscillation is reproducible when the gate voltage is swept upwards or downwards. Similar oscillations were observed in a further 50-nm-long junction that we measured, but not in any 150-nm- or 250-nm-long junctions. Therefore, we believe that the shortest junctions are ballistic while the longer ones are diffusive, allowing us to estimate the mean free path in the graphene to be between 50 nm and 150 nm. The mean free path $l$ in graphene can be precisely calculated as long as the conductivity σ and the area carrier density $n$ of the graphene are known. To determine these parameters we fabricated some additional Hall bar structures on the same CVD graphene sample, using the same fabrication technique as for the junctions. From a 4-probe measurement on the Hall bar at room temperature, we calculate that the conductivity σ of the graphene is 1.92 mS. From the Hall effect measurement at room temperature, we calculate that the area carrier density $n$ of the graphene is $5.59 \times 10^{12}$ cm$^{-2}$. (See Supporting Information for detailed measurement results.) Thus the mean free path at room temperature can be calculated as $l = \sigma h/2e^2\sqrt{n\pi} = 59$ nm. The mean free path is only expected to be a fraction higher at low temperatures, so that the value obtained from the Hall measurement is in excellent agreement with our estimated range above. The measured conductivity of graphene also implies that the contact resistance between the graphene and the superconducting electrodes accounts for a large proportion of the normal state resistance of the junction.

In conclusion, we have developed a successful method to fabricate Nb/Graphene/Nb Josephson junctions based on CVD graphene, which is important for both fundamental research and applications. The excellent quality of our junctions has been demonstrated by a series of measurements. The junctions can work over a wide temperature range from 320 mK to 1.50 K without any hysteresis, and exhibit an ideal Fraunhofer quantum interference pattern in a perpendicular magnetic field, showing the uniform distribution of the supercurrent in the junction. By carefully optimising the fabrication process, we have managed to reduce the junction length to 50 nm, shorter than the mean free path, so that evidence of ballistic transport can be observed. As far as we know, this is the first time ballistic transport has been observed in a junction based on CVD graphene. In addition, the critical current can be effectively modulated by the gate voltage, allowing more flexibility for device operation, for example simpler voltage-based control of flux or phase qubits incorporating Josephson junctions. Thanks to the relatively low cost and high quality of CVD graphene, our device is compatible with large scale production whilst maintaining its excellent properties, and so provides a competitive solution for graphene-based electronics.

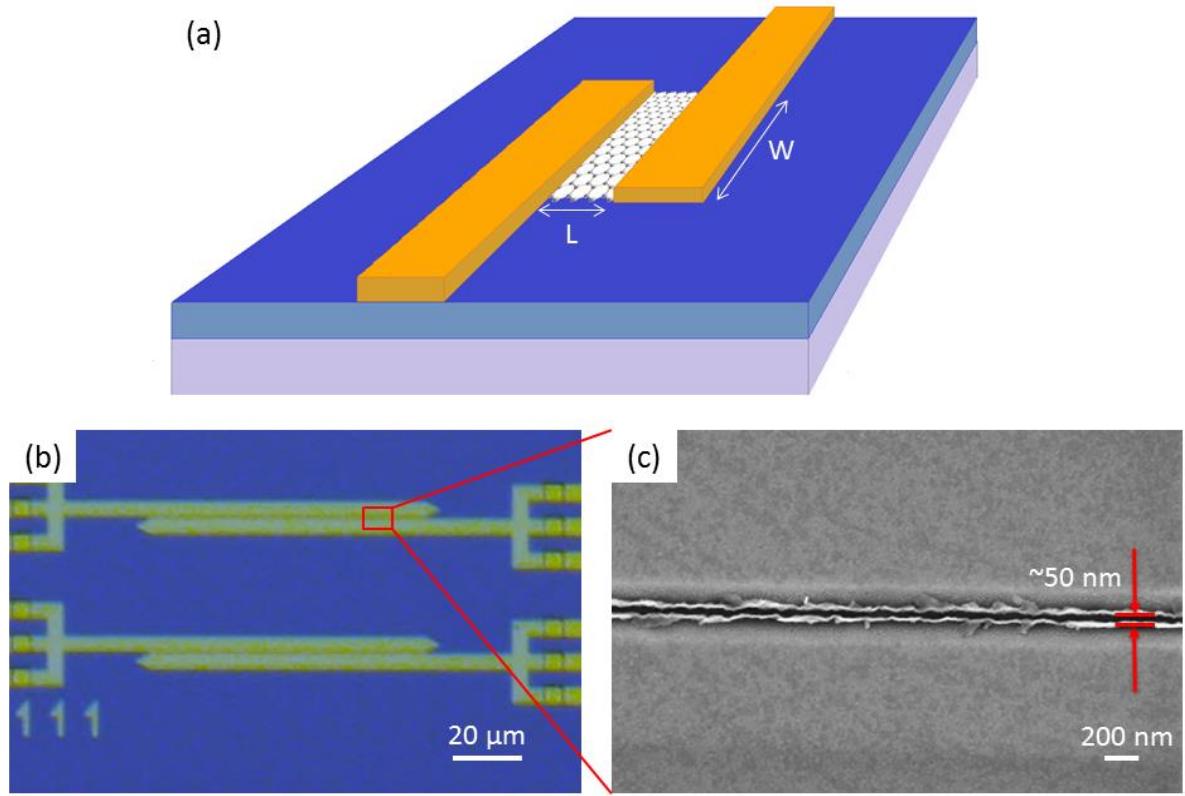

Figure 1. (a) Schematic diagram (not to scale) of a superconductor-graphene-superconductor (SGS) Josephson junction fabricated on a silicon substrate with an oxide layer. The junction is of length $L$ and width $W$. (b) Optical microscope image showing two Josephson junctions fabricated using CVD graphene and Nb electrodes. (c) SEM image of part of one junction, marked by the red box in (b), showing that the junction is only 50 nm in length.

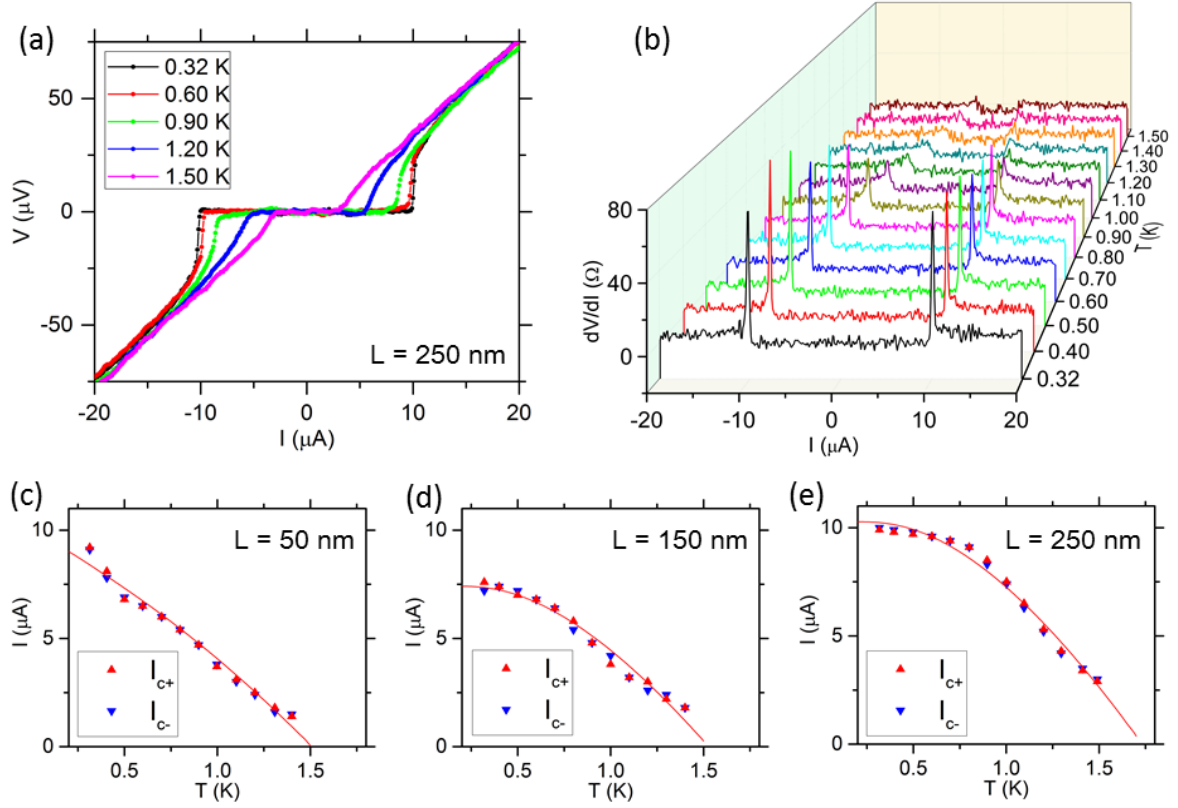

Figure 2. *I-V* characteristics of the devices versus temperature. (a) *I-V* characteristics of a 250-nm-long SGS junction measured at different temperatures. (b) The differential resistance *dV/dI* versus bias current *I*, measured on a 250-nm-long SGS junction from 0.32 K to 1.50 K. (c) – (e) The critical current $I_c$ versus temperature *T*, measured on (c) 50-nm-long, (d) 150-nm-long and (e) 250-nm-long SGS junctions, respectively. The red curve in (c) is the fitting of $I_{c+}$ by KO-2 theory, while the red curves in (d) and (e) are the fitting of $I_{c+}$ by KO-1 theory, as described in the main text.

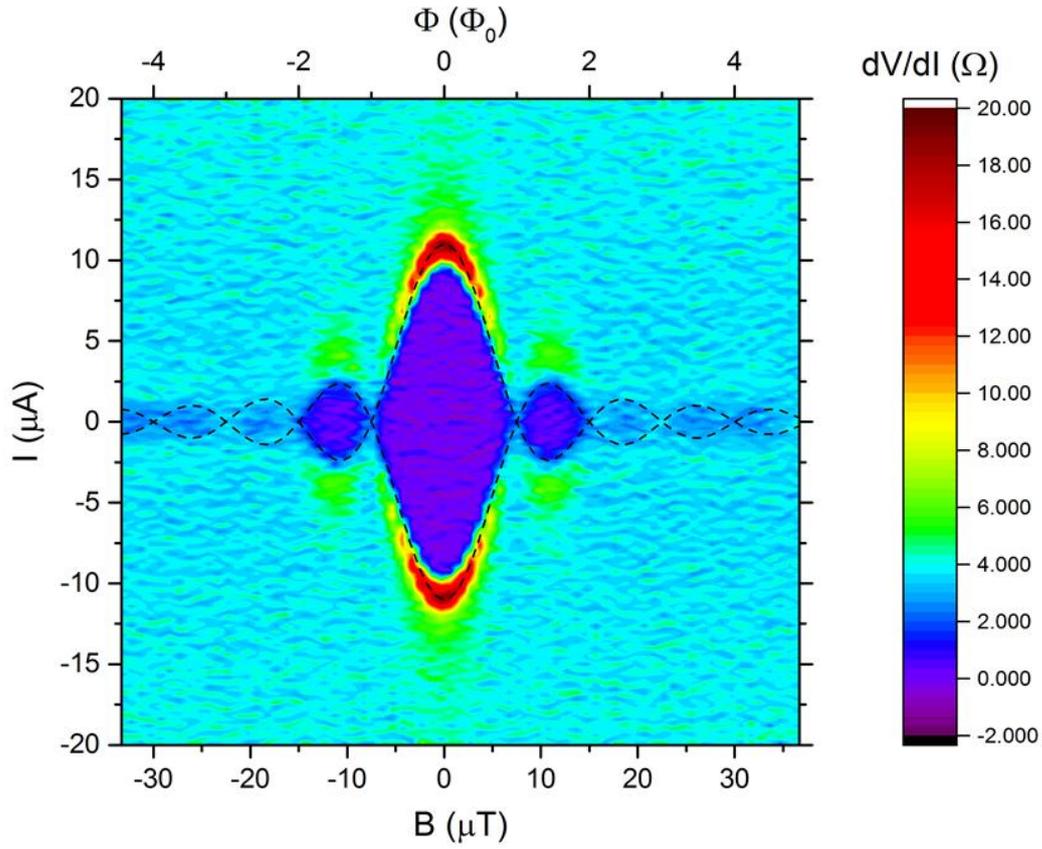

Figure 3. Colour scale plot of the differential resistance *dV/dI* as a function of bias current *I* and magnetic field *B*, measured on a 250-nm-long SGS junction at 0.32 K, showing an ideal Fraunhofer-like interference pattern. The shape of the pattern can be well fitted by $I_c \propto \sin(\pi B/\Delta B)/(B/\Delta B)$, as indicated by the black dashed lines, where $\Delta B$ is the field need to reach the first minimum.

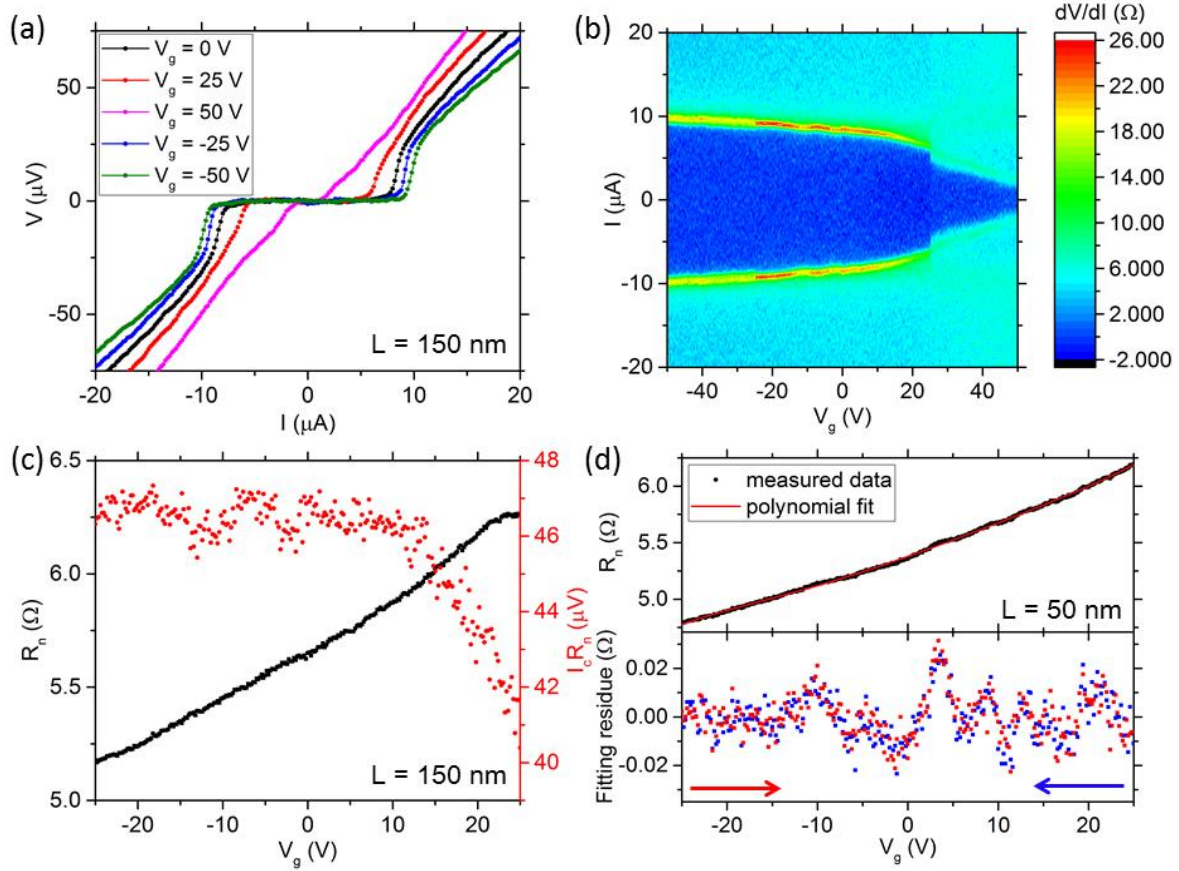

Figure 4. Electronic properties of a 150-nm-long and a 50-nm-long SGS junction versus gate voltage. (a) *I-V* characteristics of the 150-nm-long junction under different gate voltages. (b) Colour scale plot of the differential resistance *dV/dI* as a function of bias current *I* and gate voltage $V_g$, for the 150-nm-long junction. (c) Normal state resistance $R_n$ and the product $I_cR_n$ versus gate voltage $V_g$, for the 150-nm-long junction. (d) Upper panel: normal state resistance $R_n$ versus gate voltage $V_g$, for the 50-nm-long junction, fitted by a polynomial (red curve). Lower panel: fitting residue versus gate voltage $V_g$, showing periodic oscillations. The red and blue points are the residues when $V_g$ is swept upwards and downwards, respectively. All data was measured at 320 mK.